\documentclass[preprint2,tighten]{aastex6}
\pdfoutput=1 
\usepackage{amsmath,amstext}
\usepackage[T1]{fontenc}
\usepackage{apjfonts} 
\usepackage{gensymb}
\usepackage{graphicx,epstopdf}
\usepackage[figure,figure*]{hypcap}

\shorttitle{AASTeX6 Template}
\shortauthors{Byrne et al.}

\begin{document}

\title{Implementing Dust Shielding as a Criteria for Star Formation}

\author{Lindsey Byrne\altaffilmark{1}}
\email{byrnelin@u.northwestern.edu}

\author{Charlotte Christensen\altaffilmark{2}}

\author{Marios Tsekitsidis\altaffilmark{3}}

\author{Alyson Brooks\altaffilmark{4}}

\author{Tom Quinn\altaffilmark{5}}

\altaffiltext{1}{Department of Physics and Astronomy, Northwestern University, 2145 Sheridan Road, Evanston, IL 60208, United States}
\altaffiltext{2}{Physics Department, Grinnell College, 1116 Eighth Ave., Grinnell, IA 50112, United States}
\altaffiltext{3}{Department of Computer Science, Iowa State University, 2434 Osborn Dr, Ames, IA 50011, United States}
\altaffiltext{4}{Department of Physics and Astronomy, Rutgers, The State University of New Jersey, 136 Frelinghuysen Road, Piscataway, NJ 08854, United States}
\altaffiltext{5}{Department of Astronomy, University of Washington, Box 351580, Seattle, WA 98195, United States}

\begin{abstract}
Star formation is observed to be strongly correlated to dense regions of molecular gas. Although the exact nature of the link between star formation and molecular hydrogen is still unclear, some have suggested that shielding of dense gas by dust grains is the key factor enabling the presence of both. We present a sub-grid model for use in galaxy formation simulations in which star formation is linked explicitly to local dust shielding. We developed and tested our shielding and star formation models using smoothed particle hydrodynamic simulations of solar and sub-solar metallicity isolated Milky Way-mass disk galaxies. 
We compared our dust shielding-based star formation model to two other star formation recipes that used gas temperature and H$_2$ fraction as  star formation criteria. 
We further followed the evolution of a dwarf galaxy within a cosmological context using both the shielding and H$_2$-based star formation models.
We find that the shielding-based model allows for star formation at higher temperatures and lower densities than a model in which star formation is tied directly to H$_2$ abundance, as requiring H$_2$ formation leads the gas to undergo additional gravitational collapse before star formation. 
However, the resulting galaxies are very similar for both the shielding and H$_2$-based star formation models, and both models reproduce the resolved Kennicutt-Schmidt law.
Therefore, both star formation models appear viable in the context of galaxy formation simulations.

\end{abstract}


\section{Introduction}
Processes determining star formation drive the evolution of the stellar content in galaxies. 
Mimicking these processes in galaxy evolution simulations is critical to correctly modeling the stellar mass and distribution.
However, since star formation happens on scales too small to be resolved in simulations of galaxy evolution, stochastic sub-grid models must be used to model it instead.
It is therefore imperative to determine which models best mimic the actual process of star formation and produce star formation rates and locations that agree with observations from real galaxies. 
Additionally, comparing the effects of different star formation sub-grid models can reveal the primary drivers of star formation on 100 parsec scales.

Considerable observational evidence links the formation of stars to the surface density of gas in the star-forming region. \cite{Schmidt1959TheFormation.} suggested that this link takes the form of $\rho_{SFR} \propto \rho_{gas}\,^N$ and suggested that $N \approx 2$. \cite{KennicuttJr.1998TheGalaxies}  refined this relationship through observations of spiral and starburst galaxies, yielding the power law $\Sigma_{SFR} \propto \Sigma_{gas}\,^{1.4}$. However, more recent observational evidence suggests that star formation is linked specifically to the density of molecular gas. \cite{Wong2002TheGalaxies} found that star formation rate surface density is correlated more closely with the surface density of molecular hydrogen than to the total gas surface density, a result which has been replicated by others \citep{Boissier2003TheAmount, Heyer2004TheM33}. Similarly, \cite{KennicuttJr.2007StarLaw} observed a strong correlation between $\Sigma_{SFR}$ and $\Sigma_{H2}$, but no significant correlation between $\Sigma_{SFR}$ and $\Sigma_{HI}$. \cite{Bigiel2008THESCALES} also showed that the relationship between $\Sigma_{SFR}$ and $\Sigma_{gas}$ varied dramatically, while $\Sigma_{SFR}$ and $\Sigma_{H2}$ were consistently related by a Schmidt-type power law with $N = 1.0 \pm 0.2$. Various other studies have also found a strong and approximately linear relationship between $\Sigma_{SFR}$ and $\Sigma_{H2}$ \citep{Bigiel2011AGALAXIES,Schruba2011AGALAXIES,Leroy2013MOLECULARGALAXIES}, confirming the link between star formation and H$_2$, rather than total gas. 

One possible explanation for this link is that molecular hydrogen is required for star formation to occur. For example, \cite{Krumholz2009THEGAS} demonstrated that the observed relationship between galaxies' star formation rate and their atomic and molecular gas content can be explained by a model in which stars formed only out of molecular gas, at a rate given by ~1\% of the mass per free-fall time. 
Moreover, simulations in which star formation is explicitly linked to local H$_2$ abundance have produced galaxies consistent with the observed Kennicutt-Schmidt and Tully-Fisher relations \citep{Gnedin2009MODELINGSIMULATIONS,Kuhlen2012DWARFFORMATION,Christensen2012ImplementingFormation}, as well as the observed relations between ISM pressure and molecular fraction \citep{Robertson2008MolecularGalaxies}. 

However, although H$_2$ traces star formation, it need not be a direct prerequisite for star formation to occur, as cooling in dense clouds of gas occurs primarily by C$_{\text{II}}$ or CO lines. 
Indeed, by modeling star formation within a dense cloud of gas, \cite{Glover2012IsFormation} found that the presence of molecules in a gas cloud does not determine the cloud's ability to form stars.
Instead, they suggested that as long as a cloud of gas is dense and gravitationally unstable, cooling by H$_2$ is not necessary for star formation to occur. 
Therefore, it is probable that molecular hydrogen is not required for star formation but instead simply correlated with it.
For example, \cite{MacLow2012THEMODELS} have suggested that star formation and the presence of molecular gas both occur due to gravitational instability rather than being causally linked. Assuming such a situation, \cite{Krumholz2012STARGAS} demonstrated that in low-metallicity environments, below a few percent of solar metallicity, stars should be able to form without the presence of molecular hydrogen. Since the free-fall time of the gas will be longer than the thermal equilibrium timescale but shorter than the chemical equilibrium timescale, gas collapse will occur before the formation of H$_2$. This theory is supported by the high atomic, low molecular content of gamma-ray burst host galaxies, which may arise when low-metallicity gas accretes onto a galaxy, then cools and collapses into stars before H$_2$ can be formed \citep{Michaowski2015MassiveGalaxies}.

The possibility that molecular hydrogen is not necessary for star formation has implications both for our understanding of galaxy formation and for our ability to accurately model star formation in low-density and low-metallicity environments. 
In particular, the low metallicity of the high-redshift universe makes it an extreme environment for understanding star-formation.
Even after enrichment from population III stars, requiring H$_2$ for star formation could have profound effects on the high-$z$ luminosity function, especially at the low-mass end \citep{Kuhlen2012DWARFFORMATION}.
Low-H$_2$ star-forming environments also exist at low redshifts.
For example, although the outer disks of spiral galaxies lack significant amounts of molecular gas, star formation can still occur, albeit inefficiently \citep{Ferguson1998DiscoveryGalaxies, Thilker2006,Bigiel2010EXTREMELYGALAXIES}. The same is true for the low-surface density, low-metallicity environments of dwarf galaxies  \citep[e.g.][]{vanZee1997AGalaxies, Leroy2006Molecular10, Melena2009BRIGHTGALAXIES, Roychowdhury2009StarGalaxies}.  
In order to accurately model these environments, we must understand the process of star formation in environments with low amounts of H$_2$. 

If molecular hydrogen is not a prerequisite for star formation, other factors must be considered.
One potential explanation is that shielding by dust leads to both the formation of molecular hydrogen and the formation of cold gas necessary for gravitational collapse and star formation. \cite{Schaye2004StarWhere} argued that the transition of gas to the cold phase at high column densities both allows star formation to occur and coincides with an increase in the molecular fraction, as H$_2$ forms more readily in dense gas and contributes to further cooling.
Using detailed simulations of chemical evolution within collapsing gas clouds, \cite{Glover2012IsFormation} found that the column density of gas clouds and the associated dust shielding are the determining factors for both the creation of molecules and star formation. Dust shielding both reduces photodissociation of H$_2$ by Lyman-Werner (LW) radiation and blocks photoheating by the Interstellar Radiation Field (ISRF); this allows cold, dense gas to form in the same regions where molecular hydrogen is able to form and remain. 
They argue that shielding of the ISRF, not the presence of H$_2$ or CO, is the most important factor enabling star formation in gas clouds, a conclusion which has also been reached by others \citep[e.g.][]{Krumholz2011WHICHRATE,Clark2014OnRate}. 

The evidence for a strong connection between dust shielding and star formation suggests that the accuracy and realism of numerical simulations may be increased by making star formation explicitly dependent on dust shielding. Such a model might allow for star formation to occur in areas or on timescales that a H$_2$-based recipe would preclude. Here we implement a dust shielding-based star formation recipe and compare ISM and star formation properties of the resulting simulated galaxies to those produced using a temperature ceiling model and a H$_2$-based model. We further analyze the effects of shielding versus H$_2$-based star formation on a cosmological simulation of a dwarf galaxy by comparing disk size, star formation history, and the location of star formation. 
The comparison of shielding to other criteria enables a greater understanding of the link between dust shielding and star formation, and therefore how to more accurately model galaxy formation.

\section{Methodology}
We implement a novel method for dust shielding-based star formation for use within smoothed particle hydrodynamic galaxy formation simulations, including incorporating an updated method for calculating shielding.
Below, we describe the isolated disk galaxy and cosmological galaxy formation simulations used to test our star formation and shielding models (\S\ref{sec:ic}).
We then introduce the smoothed particle hydrodynamic code used (\S\ref{sec:code}) and describe our implementation of shielding (\S\ref{sec:shield}) and star formation (\S\ref{sec:sf}).

\subsection{Initial conditions}\label{sec:ic} 
To test our shielding and star formation models, we used simulations of a $1.4\times 10^{12}$ M$_\odot$ isolated Milky Way-like disk galaxy within a \citet{Navarro1997} dark matter halo. 
Following the work of \citet{Kaufmann2007AngularFormation} and \citet{Stinson2007BreathingFormation}, disks were formed by allowing a cloud of virialized gas to collapse within a virialized dark matter halo with a concentration of $c = 8$, where $c = R_{vir}/R_s$, and $R_s$ is the scale radius of the Navarro-Frenk-White (NFW) profile.
The initial halo had a gas fraction of 10\%, and the gas particles were given an initially uniform circular velocity such that the dimensionless spin parameter $\lambda = (j_{gas}|E|^{1/2})/(GM^{3/2}) = 0.039$, where $j_{gas}$ is the average specific angular velocity of the gas, and E and M are the total energy and mass of the halo. 
The initial masses of the star, gas, and dark matter particles were $3.4\times 10^4$ M$_\odot$, $1.4\times 10^5$ M$_\odot$, and $1.3\times 10^6$ M$_\odot$ respectively. 
The gravitational softening length was 206 pc and the smoothing lengths had a minimum value of 0.1 times the softening length.
After 1 Gyr, following the formation of a stable disk,
we scaled the gas metallicities to create one simulation with average ISM metallicity of 1.0 Z$_\odot$ and one with 0.1 Z$_\odot$.  
These two isolated disk galaxies formed the initial conditions for testing our star formation and shielding models.
We then enabled H$_2$ physics along with our various shielding and star formation models described below, and integrated the simulations for an additional 1.0 Gyr.

We also tested the effects of our shielding implementation using a zoom-in \citep{Katz93} cosmological simulation of a dwarf galaxy, $M_{vir} = 4\times 10^{10}$ M$_\odot$. 
A dwarf galaxy was chosen to highlight the impact of our model in a low-metallicity environment. 
The initial conditions consisted of a 25-Mpc box selected to surround a halo chosen from a low-resolution, dark-matter-only simulation. The {\sc cmbfast} code was used to calculate the initial power spectrum used for the linear density field. We assumed a $\Lambda$CDM cosmology with values from the \textit{Wilkinson Microwave Anisotropy probe}
\citep{Spergel2003FirstYearParameters}. 
The masses of the star, gas, and dark matter particles were 1000, 3000, and 16,000 solar masses respectively, and the softening lengths were 87 pc. 

The initial conditions for this simulation were first used in \cite{Governato2010BulgelessOutflows} and produced a bulgeless dwarf galaxy. \cite{Christensen2012ImplementingFormation} simulated the same galaxy with the addition of metal line cooling and H$_2$-based star formation. 
Compared to a simulation with metal-line cooling and density-based star formation, they found that the addition of H$_2$-based star formation produced a clumpier ISM and on-going star formation at $z = 0$. 

\subsection{General description of code}\label{sec:code}
The simulations were performed using the galaxy formation code {\sc ChaNGa}, a N-body smoothed particle hydrodynamics (SPH) code \citep{N-BodyShop2011ChaNGa:Solver, Menon2015}.
{\sc ChaNGa} is a highly scalable successor to {\sc gasoline} \citep{Wadsley2004Gasoline:TreeSPH}, itself an extension of the parallel, gravity tree code PKDGRAV \citep{Stadel2002CosmologicalAnalysis}. 
{\sc ChaNGa} employs a modern implementation of the SPH kernel that uses a geometric mean density in the force expression \citep{Ritchie2001, Menon2015, Wadsley2017Gasoline2:Code}, thus eliminating artificial gas surface densities.

{\sc ChaNGa} incorporates a suite of physical modules for thermodynamics, star formation, and stellar feedback, as described below. 
Non-equilibrium ion abundances, including H$_2$, are calculated by integrating over the H and He chemical networks. Heating and photoionization are determined using a cosmic ultraviolet background (Haardt \& Madau 2005)\footnote{ Haardt \& Madau (2005) refers to an unpublished updated version of \citet{Haardt1996RadiativeBackground}, specified in {\sc cloudy} \citep{Ferland98} as ``table HM05.''}. 
Cooling is calculated via collisional ionization \citep{Abel1997ModelingCosmology}, radiative recombination \citep{Black1981TheClouds,Verner1996AtomicTemperature},  photoionization, bremsstrahlung, and H, He, and metal line cooling \citep{Cen1992AMethodology, Shen2010TheDiffusion}. To calculate the non-equilibrium abundance of molecular hydrogen, we adopt the model used by \cite{Christensen2012ImplementingFormation}. Local, non-equilibrium H$_2$ abundances are determined by the rates of dust grain H$_2$ formation, gas-phase H$_2$ formation, collisional dissociation, and photodissociation by Lyman-Werner radiation from nearby stellar particles. HI and H$_2$ are both shielded by dust and H$_2$ is additionally self-shielded. The Lyman-Werner flux used for calculating the rate of H$_2$ photodissociation and heating for a given gas particle is approximated based on the cosmic ultraviolet background and the average Lyman-Werner flux from nearby star particles. In calculating the stellar Lyman-Werner radiation, the particle gravity tree is used to select proximate gas particles, as described in \cite{Christensen2012ImplementingFormation}. This method avoids a full radiative-transfer calculation while still approximating the spatial and temporal variation in the Lyman-Werner flux. When determining the rates of H$_2$ photodissociation and heating, we further include a sub-grid model for dust and self-shielding. This same dust shielding model is also used in determining the rates of photoionization and heating of HI by the cosmic ultraviolet background. The sub-grid shielding models and the different techniques we explored to estimate column length are described in detail in \S 2.3. Star formation rates are determined stochastically based on local gas properties, described for each model in section 2.4. 

The simulations use the blastwave supernova (SN) feedback model described in \cite{Stinson2006StarGalaxies}, in which each SN releases $10^{51}$ ergs of energy into the surrounding gas.
For SNe II, radiative cooling is disabled within the blastwave radius during the theoretical lifetime of the momentum-conserving phase of the supernova remnant. This sub-grid recipe prevents artificial cooling and allows affected gas to naturally rise from the disk without the use of a momentum kick.
SNe I and II also inject metals into the ISM.
These simulations assume a \citet{Kroupa2001} initial mass function, SNe rates from \citet{Raiteri96}, and metal yields from \citet{Woosley1995} and \citet{Thielemann1986}.
Metals are also returned to the ISM by stellar winds using mass loss rates from \citet{Weidemann87}, and they are dispersed using a shear-dependent subgrid turbulent mixing model \citep{Shen10} with a metal diffusion coefficient of 0.03.

\subsection{Shielding recipes}\label{sec:shield}
Shielding is responsible for the preservation of H$_2$ in dense gas as both dust and self-shielding protect the gas from photodissociation.
We examine the role that dust shielding may play in promoting star formation by preventing photoheating.
As part of these efforts, we improve our shielding models through the implementation of a new model for calculating column length based on \citet{Safranek-Shrader2017ChemistryDiscs}.

Shielding has significant effects on scales below the resolution limit of our simulations. 
We account for these effects with a local, sub-grid approximation of shielding, following the work of \cite{Draine1996StructureFronts}, \cite{Glover2007SimulatingConditions}, and \cite{Gnedin2009MODELINGSIMULATIONS}. The dust shielding $S_d$ and self-shielding $S_{H_2}$ functions were implemented as in \cite{Christensen2012ImplementingFormation}: 
We calculated the dust shielding as
\begin{equation}
S_d = e^{-\sigma_{d_\text{eff}} Z/Z_{\odot}(N_{HI}+2N_{H2})}
\label{eq:s_dust}
\end{equation}
where N$_{\text{HI}}$ and N$_{\text{H2}}$ are the column densities of H$_\text{I}$ and H$_2$ respectively, $Z$ is the metallicity, and $\sigma_{d_\text{eff}}$ is a parameter representing the effective attenuation cross section, tuned to be $2\times 10^{-21}$ cm$^2$. 
The H$_2$ self shielding is calculated as
\begin{equation}
S_{H_2} = \frac{1-\omega_{H_2}}{(1+x^2)}+\frac{\omega_{H_2}}{(1+x)^{(1/2)}}e^{-0.00085(1+x)^{1/2}}
\end{equation}
where $\omega_{H_2}$ is an adjustable parameter tuned to be 0.2 and $x = N_{H2}/(5\times 10^{14})$ cm$^2$.

\begin{figure}[b]
\begin{center}
\includegraphics[width=0.5\textwidth]{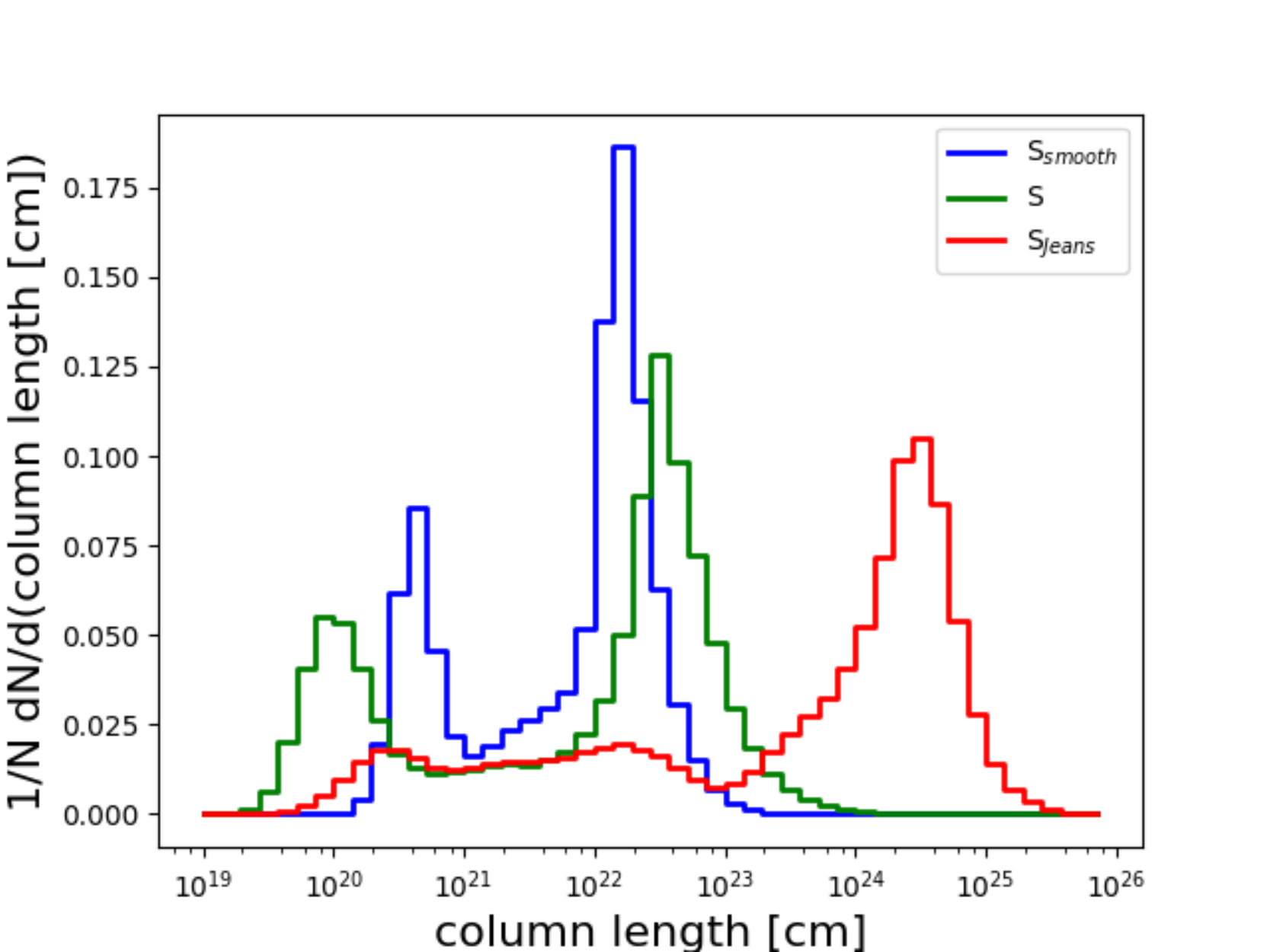}
\end{center}
\caption{Histogram of column lengths for the particles in the three shielding simulations. Column lengths are calculated as in the respective shielding models: as the smoothing length for S$_{smooth}$, the temperature-capped Jeans length for S, and the Jeans length for S$_{Jeans}$. Column lengths are artificially high in the S$_{Jeans}$ simulation due to the temperature dependence of the Jeans length.}
\label{fig:1}
\end{figure}

The column densities are the product of the column length associated with each gas particle and the number densities of HI and H$_2$.
The molecular hydrogen recipe introduced in \cite{Christensen2012ImplementingFormation} assumed the column length to be equal to the particle smoothing length, $h$, a recipe which we will refer to as S$_{smooth}$.
However, this recipe has the potential to be susceptible to resolution effects. 
More recently, \cite{Safranek-Shrader2017ChemistryDiscs} evaluated several common approximations for shielding length---including the Sobolev approximation, the Jeans length, a length based on the local density and its gradient, a power-law approximation, and a single-cell approximation---and compared them to a detailed ray-tracing solution to the radiative transfer problem.
They found that a temperature-capped Jeans length performed well at matching the effective visual extinction and performed better than other local models at calculating mass-weighted abundances of H$_2$ and CO. 

In the S model, we consider the shielding length to be equal to the particle Jeans length using the empirically-chosen temperature ceiling of 40 K from \cite{Safranek-Shrader2017ChemistryDiscs}: 
\begin{equation}
L_{shield} = \sqrt{\frac{15k_BT}{4 \pi G m_H \rho}}
\end{equation}
where T is the minimum of the particle's temperature and 40 K, $m_H$ is the mass of atomic hydrogen, and $\rho$ is the gas density. 
We also investigated the use of a Jeans length approximation for shielding length without a temperature ceiling (the S$_{Jeans}$ model).

In order to assess the differences among our shielding recipes, we examined the resulting column lengths of gas particles in isolated disk galaxies generated using the S$_{smooth}$, S, and S$_{Jeans}$ models. We found that the S$_{Jeans}$ model resulted in significantly longer column lengths than the S$_{smooth}$ or S models (Figure~\ref{fig:1}). 
In that model, high-temperature, low-density gas is modeled as having a long column length, increasing the column densities and dust shielding beyond what would be expected  in reality. This supports the findings of \cite{Safranek-Shrader2017ChemistryDiscs}, who concluded that H$_2$ and CO abundances were significantly over-predicted by a Jeans length shielding model without a temperature cap. The temperature-capped Jeans length, however, led to column lengths similar to the smoothing lengths in the S$_{smooth}$ simulation. 

\begin{figure}[t]
\begin{center}
\includegraphics[width=0.5\textwidth]{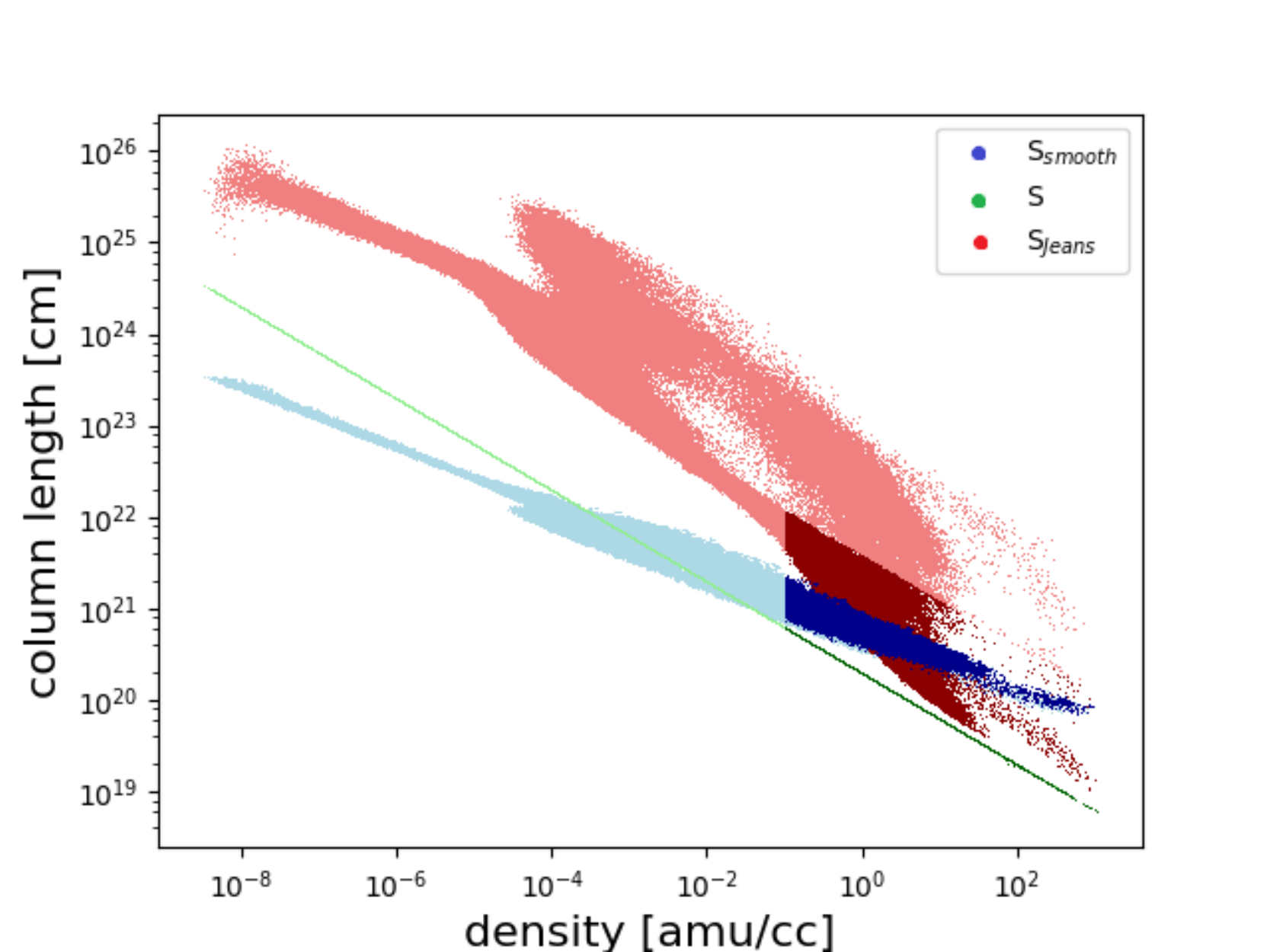}
\end{center}
\caption{Column length vs. density for the particles in the three shielding simulations. Dark colors indicate particles that meet the temperature and density requirements for star formation, i.e. $\rho \geq .1$ amu cm$^{-3}$ and T $\leq 15000$ K. Imposing a temperature ceiling on the Jeans length calculations leads to significantly lower column lengths overall in the S simulation than the S$_{Jeans}$ simulation, though the discrepancy is smaller for potentially star-forming gas. The S column lengths are higher than the S$_{smooth}$ column lengths at low densities, but lower at the densities at which star formation is possible.}
\label{fig:2}
\end{figure}

The extent of the impact of the different column length recipes is further analyzed in Figure~\ref{fig:2}, which compares column lengths as a function of density. Although the column lengths in the S$_{Jeans}$ simulation are significantly longer on average than the column lengths in the other two shielding simulations, the effect is less significant when looking only at gas particles eligible for star formation, indicated by darker shades in the diagram. The S model creates the shortest column lengths at high (i.e. potentially star-forming) densities. 
We also note that since almost all gas has temperatures greater than 40 K, the S column lengths are generally simply proportional to $\rho^{1/2}$. Additionally, for gas particles with equal masses and T > 40 K, the column lengths for S and S$_{smooth}$ will be correlated with each other.

\begin{figure}[t]
\begin{center}
\includegraphics[width=0.5\textwidth]{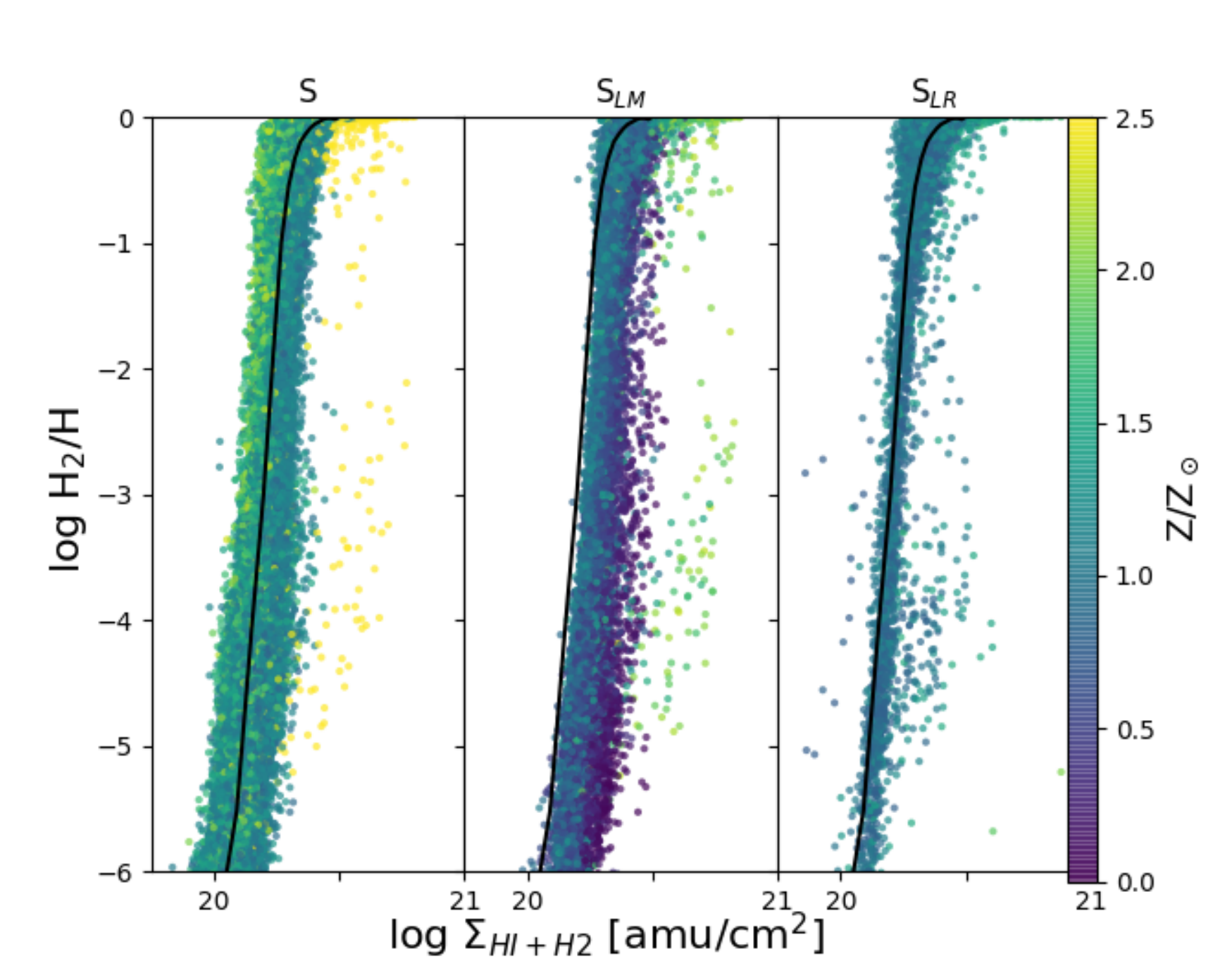}
\end{center}
\caption{The transition from atomic to molecular hydrogen as a function of surface density using the S model for an isolated, solar metallicity Milky Way-like simulation (left) compared to a low-metallicity ($\langle Z_{ISM} \rangle =0.1 \; Z_\odot$ for the initial conditions) simulation (center) and a low-resolution solar metallicity simulation (right) using the same star formation model. The points represent individual SPH gas particles colored according to their metallicity. The black line represents the median H$_2$ fraction at a given surface density for the S model. Surface densities were calculated by multiplying the HI and H$_2$ densities of each particle by its temperature-capped Jeans length. As metallicity decreased, the transition moves to slightly higher column densities since dust shielding and the rate of H$_2$ formation decreases. The transition occurs at the same density for the high and low resolution simulations, indicating that the transition is independent of the number of particles over this range of resolutions. 
}
\label{fig:3}
\end{figure}

To further test our shielding model, we examined the HI+H$_2$ surface densities at which hydrogen transitioned from atomic to molecular for simulations using the S model at sub-solar metallicity (S$_{\mathrm{LM}}$) and with a factor of ten lower mass resolution (S$_{\mathrm{LR}}$) (Figure~\ref{fig:3}). 
Surface densities were calculated per-particle as $\Sigma_{HI+H2} = \rho  (N_{HI}+2N_{H2}) L_{shield}$.
As metallicity decreased, the transition moved to slightly higher column densities since dust shielding, and therefore the rate of H$_2$ formation, decreased. 
While the metallicity of the initial conditions differed by a factor of 10, the shift in the transition was roughly a factor of 2. 
The reduced shift was because of a combination of factors.
First, self-shielding, which is not metallicity-dependent, is significant at high densities, enabling the transition from HI to H$_2$ to occur at lower surface densities than dust shielding alone would predict.
Secondly, although the average metallicity was initially $z=0.1 \; Z_\odot$ for the low-metallicity simulation, we found that the difference in metallicities between the two simulations had decreased after they were integrated, particularly for the molecular gas. 
The average ISM metallicities of each of the simulations after being integrated were $\langle Z_{ISM} \rangle= 1.3\; Z_\odot$ for S, $\langle Z_{ISM} \rangle= .73\; Z_\odot$ for S$_{\mathrm{LM}}$, and $\langle Z_{ISM} \rangle = 1.0\; Z_\odot$ for S$_{\mathrm{LR}}$.

The low-resolution simulation was generated from a similar set of initial conditions as the other isolated, Milky Way-like simulations.
The only differences were a factor of 10 lower mass resolution and $10^{1/3}$ lower spatial resolution.
Specifically, the low-resolution simulation had $10^5$ gas particles with initial masses $1.4\times 10^6$ M$_\odot$, instead of $10^6$ gas particles with initial masses $1.4\times 10^5$ M$_\odot$. 
Decreasing the mass resolution of the simulation did not significantly affect the surface density at which the transition occurred, indicating that that the transition is consistent over this range of mass resolution. 

\subsection{Star formation models}\label{sec:sf}
We compared three star formation models to determine the impact of tying star formation to H$_2$, shielded gas, or cold gas (Table~\ref{tab:1}). 
Our reference model (H2) was the H$_2$-based star formation model outlined in \citet{Christensen2012ImplementingFormation}.
We compare that model to one in which the star formation probability depends on the amount of dust shielding (S).
We also examine the case in which the star formation is limited to cold gas (TC), as the motivation for shielding-dependent star formation is that shielding enables gas to cool to low temperatures.

\begin{table}[t]
\begin{center}
\begin{tabular}{|c|c|c|c|c|} \hline
	Model & SF recipe & L$_s$ & T$_{max}$ [K] & $c^\ast$ \\ \hline
    \textbf{TC} & Temperature cap & $\lambda_J$ (T $\leq 40$) & 1000 & $c_0^\ast$ \\
    \textbf{H2} & H$_2$ fraction & $\lambda_J$ (T $\leq 40$) & 15000 & $c_0^\ast (X_{H_2})$ \\
	\textbf{S} & Shielding fraction & $\lambda_J$ (T $\leq 40$) & 15000 & $c_0^\ast (1-S_d)$ \\
    \hline
\end{tabular}
\end{center}
\caption{Star formation and H$_2$ formation parameters of the models analyzed in this paper. L$_s$ is the shielding length used to calculate H$_2$ abundance and (if applicable) star formation. $\lambda_J$ is the Jeans length of a gas particle, while $h$ is the smoothing length. T$_{max}$ is the maximum temperature at which gas particles can form stars. $c^\ast$ is the star formation efficiency factor. For all models, the minimum density at which stars can form is $\rho_{min} = .1$ amu cm$^{-3}$.} 
\label{tab:1}
\end{table}

The probability of a sufficiently cold and dense gas particle spawning a star particle is adopted from \cite{Stinson2006StarGalaxies}: 
\begin{equation}
p = \frac{m_{gas}}{m_{star}}(1 - e^{-c^\ast \Delta t/t_\text{form}}),
\end{equation}
where $m_{gas}$ and $m_{star}$ are the masses of the gas particle and the potential star particle respectively, \(t_{form}\) is the local dynamical time, and \(c^\ast\) is a star-forming efficiency factor, whose functionality depends on the model, as described below. 

In the H2 recipe, star-forming efficiency was linked to the local abundance of molecular hydrogen: 
\begin{equation}
c^\ast = c_0^\ast X_{H_2}
\end{equation}
where $X_{H_2}$ is the number fraction of H$_2$, and \(c_0^\ast\) is an adjustable parameter tuned to 0.1  for the isolated galaxies and 0.01 for the cosmological simulations, values chosen to produce reasonable Kennicutt-Schmidt relations for the H2 model. 
In the shielding recipes, we modified the efficiency factor to be dependent on dust shielding: 
\begin{equation}
c^\ast = c_0^\ast(1-S_d)
\end{equation}
where $S_d$, as defined in equation~\ref{eq:s_dust}, is the fractional reduction in the intensity (so $S_d$ = 1 corresponds to no shielding by dust).
We used the same values of $c_0^\ast$ as in the H$_2$ recipe.
In the temperature-based recipe (TC), only those particles passing a more stringent temperature cut were allowed to form stars, and we used a constant value of $c^\ast = c_0^\ast$. 
We only tested the TC recipe on the isolated galaxies and used the same value of $c_0^\ast = 0.1$ as in the H2 and S cases for those galaxies.

All models required gas particles to meet temperature and density criteria ($\rho \geq \rho_{min}$ and $T \leq T_{max}$) to be eligible for star formation. 
The same density threshold was used across all models: $\rho_{min} \geq 0.1$ amu cm$^{-3}$, a value chosen to be low enough to have no significant effect on star formation. 
For H$_2$ and shielding-based star formation models, $T_{max}$ was set to 15000 K, a temperature cut high enough to have no significant effect on star formation.
For TC, $T_{max}$ was lowered to 1000 K, a temperature chosen to be similar to the upper temperature at which particles in the simulation contained significant amounts of H$_2$.
This temperature ceiling must be higher than the observed temperature of molecular clouds to reflect the limited resolution of the simulations.

\begin{figure*}[t]
\begin{center}
\includegraphics[width=\textwidth]{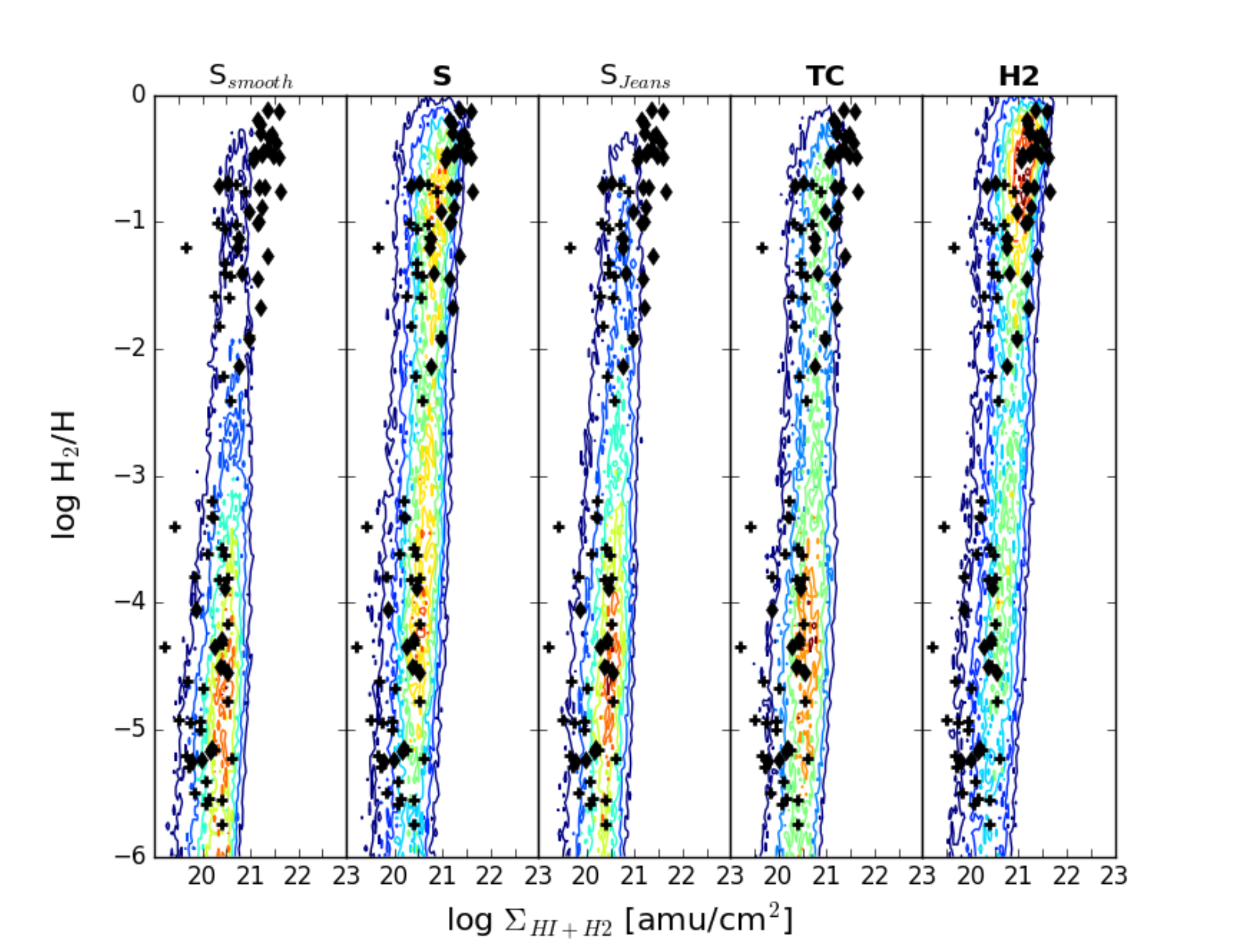}
\end{center}
\vspace{-5pt}
\caption{The transition from atomic to molecular hydrogen as a function of surface density for isolated, Milky Way-like simulations with five different models.
The diamonds are observational data from \citet{Wolfire2008ChemicalFormation} and the plus signs are observational data from \citet{Gillmon2006AAGNs}. 
The linearly spaced contours were generated from mock observations of the simulations. 
They represent the distribution of surface densities and H$_2$ fractions from individual cells in a velocity cube.
The first three panels compare the results of shielding-based star formation with three different methods for calculating shielding -- our preferred S model, the original S$_{smooth}$ model, and a non-temperature capped Jeans length model, S$_{Jeans}$.
The final two panels show the S shielding model with two different star formation models, a H$_2$-based model (H2), and a temperature capped model (TC).
All models produce HI-to-H$_2$ transitions at appropriate surface densities; however, different star formation criteria allow for the presence of varying amounts of H$_2$. 
The S and H2 simulations most successfully replicate the  distribution of the H$_2$ fraction seen in the observational data.
In contrast, the S$_{smooth}$, S$_{Jeans}$, and TC models result in only small amounts of high surface density material.
}
\label{fig:5}
\end{figure*}

In order to test the accuracy of our models, we compared the HI+H$_2$ surface densities at which hydrogen transitioned from atomic to molecular for isolated disk simulations at solar metallicity. 
The surface densities are plotted in Figure~\ref{fig:5} and compared to Milky Way observational data from \cite{Gillmon2006AAGNs} and \cite{Wolfire2008ChemicalFormation} determined from absorption spectroscopy. 
In addition to showing the results of the three different star formation models, we also include shielding-based star formation using the previously discussed S$_{smooth}$ and S$_{Jeans}$ shielding models. 
In all other simulations, shielding was calculated using the S model. 

In order to ensure an accurate comparison to observations, the surface densities and H$_2$ fractions were calculated from mock observations generated by post-processing the galaxy to create velocity cubes of HI and H$_2$. The cubes were produced by using the smoothing kernel to spatially distribute the gas particles, and then calculating the expected amount of 21-cm emissions and binning it in x, y, and velocity space. Data was generated in 128 velocity channels with 1.3 km s$^{-1}$ bins and a 5 pc pixel size. 

All models were able to successfully reproduce the observed transition from atomic to molecular hydrogen, indicating that all shielding models produced reasonable behavior. 
The different star formation criteria allowed for different amounts of H$_2$, as will be discussed further in \S 3. 
The H2 and S models produced the greatest amounts of H$_2$-rich gas, as in the other models, gas particles frequently formed stars before a large amount of H$_2$ could accumulate.

\begin{figure}
\begin{center}
\includegraphics[width=0.5\textwidth]{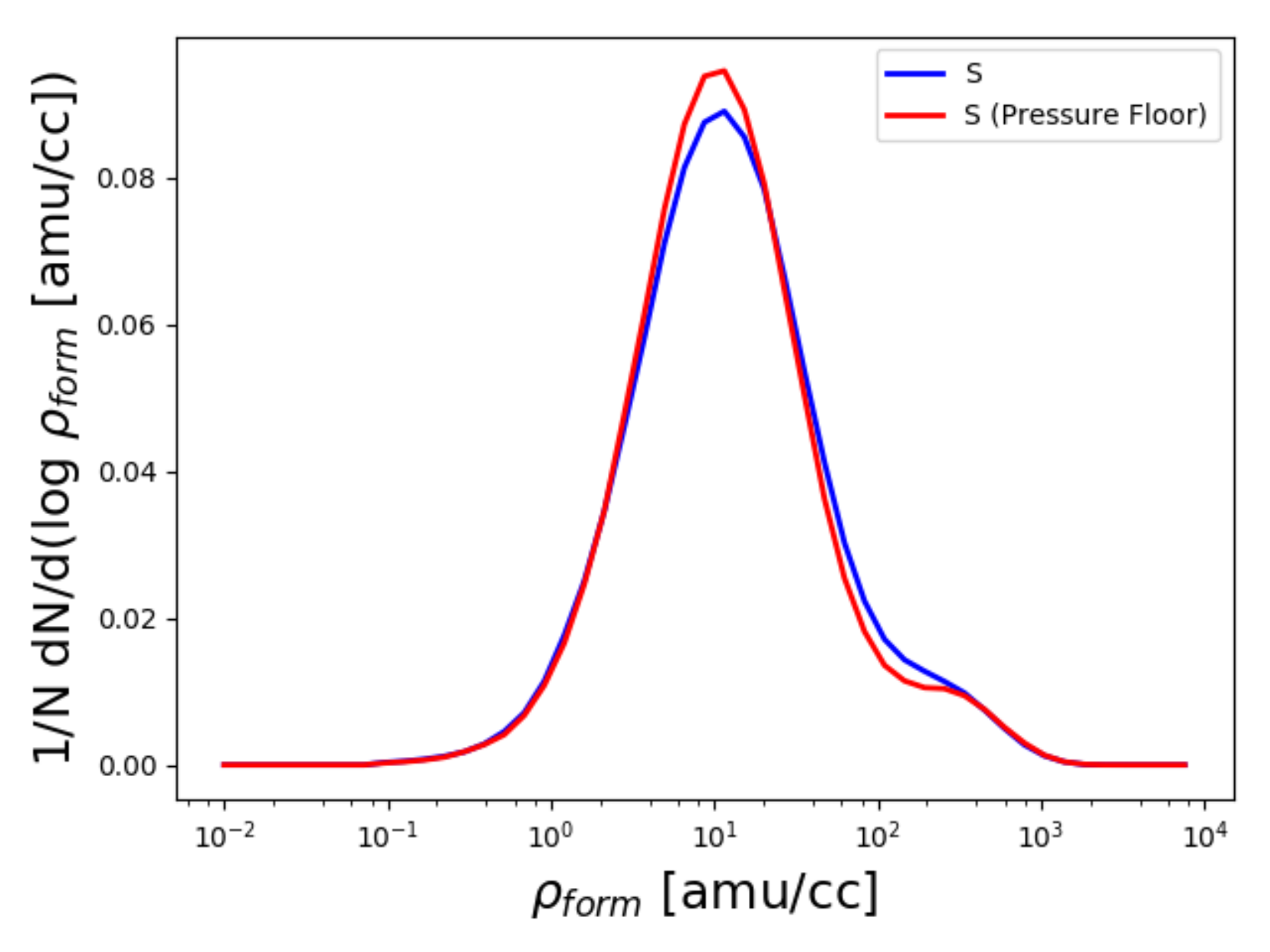}
\end{center}
\caption{Normalized histograms of the density of star-forming gas for a galaxy formed with the S shielding model, with and without the addition of a Jeans pressure floor. The presence of a pressure floor did not significantly affect the density at which gas formed into stars.}
\label{fig:4}
\end{figure}

Linking star formation to molecular or shielded gas relegates star formation to generally dense and cold gas for which the Jeans mass may no longer be resolved.
\citet{Robertson2008MolecularGalaxies}, therefore, advocated the adoption of a pressure floor to prevent numerical Jeans fragmentation of dense, cold gas.
However, the use of a pressure floor has also been rejected as artificial or unnecessary.
For example, \cite{Christensen2012ImplementingFormation} found that the addition of a Jeans pressure floor did not significantly affect the number or distribution of stars formed for simulations of this resolution, as gas particles for which the Jeans mass and length are unresolved necessarily meet the star formation criteria. 
\citet{Hopkins2017} similarly found artificial pressure floors in SPH simulations to be unnecessary and unphysical.
We tested the impact of adding a Jeans pressure floor to the S shielding model and found that, due to the high resolution of the simulation, it did not create any significant differences in star formation history, gas density, or gas distribution. The average density at which gas formed into stars was not affected by the presence of a Jeans floor, as shown in Figure~\ref{fig:4}. All models discussed in this paper therefore do not have an artificial pressure floor.

\section{Results}

We analyzed the effects of including shielding-based star formation and compared it to the more traditional star formation recipes of a temperature-capped model and a H$_2$ abundance-based model, as shown in \S\ref{sec:results_iso}. 
We used solar and sub-solar metallicity isolated disk simulations to examine the properties of the star-forming gas for each of the three models.
By examining the temperatures and densities at which gas formed into stars, we were able to assess how the different star formation recipes affected the environments within which star formation occurred.

In \S\ref{sec:results_cosmo}, we further analyzed the effect of using H$_2$ vs shielding-based star formation on simulations of more realistic galaxies by comparing the evolution of two cosmological dwarf galaxy simulations.
These simulations were chosen to highlight the effects of the models in low-metallicity environments.
We focused our analysis on the star formation within the simulations, and compared the star formation histories, Kennicutt-Schmidt relation, and stellar and gas density profiles for the simulations.
Due to the computational expense, we limited our analysis of cosmological simulations to the H2 and S models.

\subsection{Star-Forming Gas in Isolated Simulations}
\label{sec:results_iso}
\begin{figure}[t]
\begin{center}
\includegraphics[width=.5\textwidth]{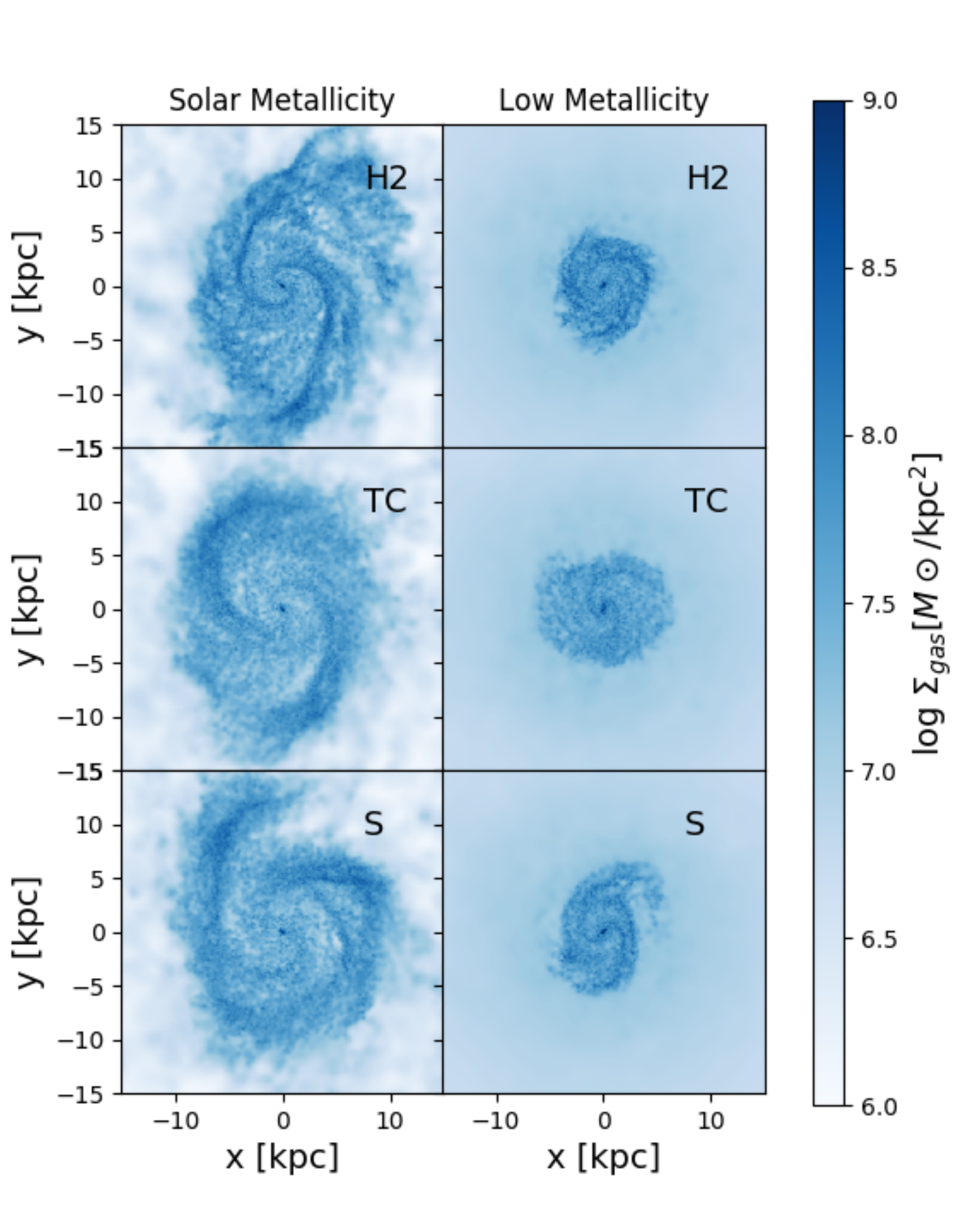}
\end{center}
\vspace{-10pt}
\caption{Gas surface density after 1 Gyr for isolated disk galaxies at solar (left column) and sub-solar metallicities (right column), simulated with the H2 (top panels), TC (middle panels) and S (bottom panels) star formation criteria. 
The disks of the H2 and S simulations were denser than those of the TC model, resulting in more pronounced spiral structure.}
\label{fig:6}
\end{figure} 

Figure~\ref{fig:6} illustrates the ISM for solar and sub-solar isolated disk galaxies simulated with the TC, H2, and S star formation models.
The star formation criteria has a clear effect on the structure of the gas. 
The H2 and S simulations led to formation of denser gas than the TC model. 
This effect is especially pronounced in the H2 model at solar metallicity, in which the greater amount of dense and cold gas results in the formation of more pronounced spiral arms.
For all star formation models, lower metallicity gas resulted in more compact disks, because the reduced cooling efficiency prevented gas from reaching high densities, and low-metallicity gas took longer to collapse into a disk for a given initial radius.

\begin{figure}
\begin{center}
\includegraphics[width=.5\textwidth]{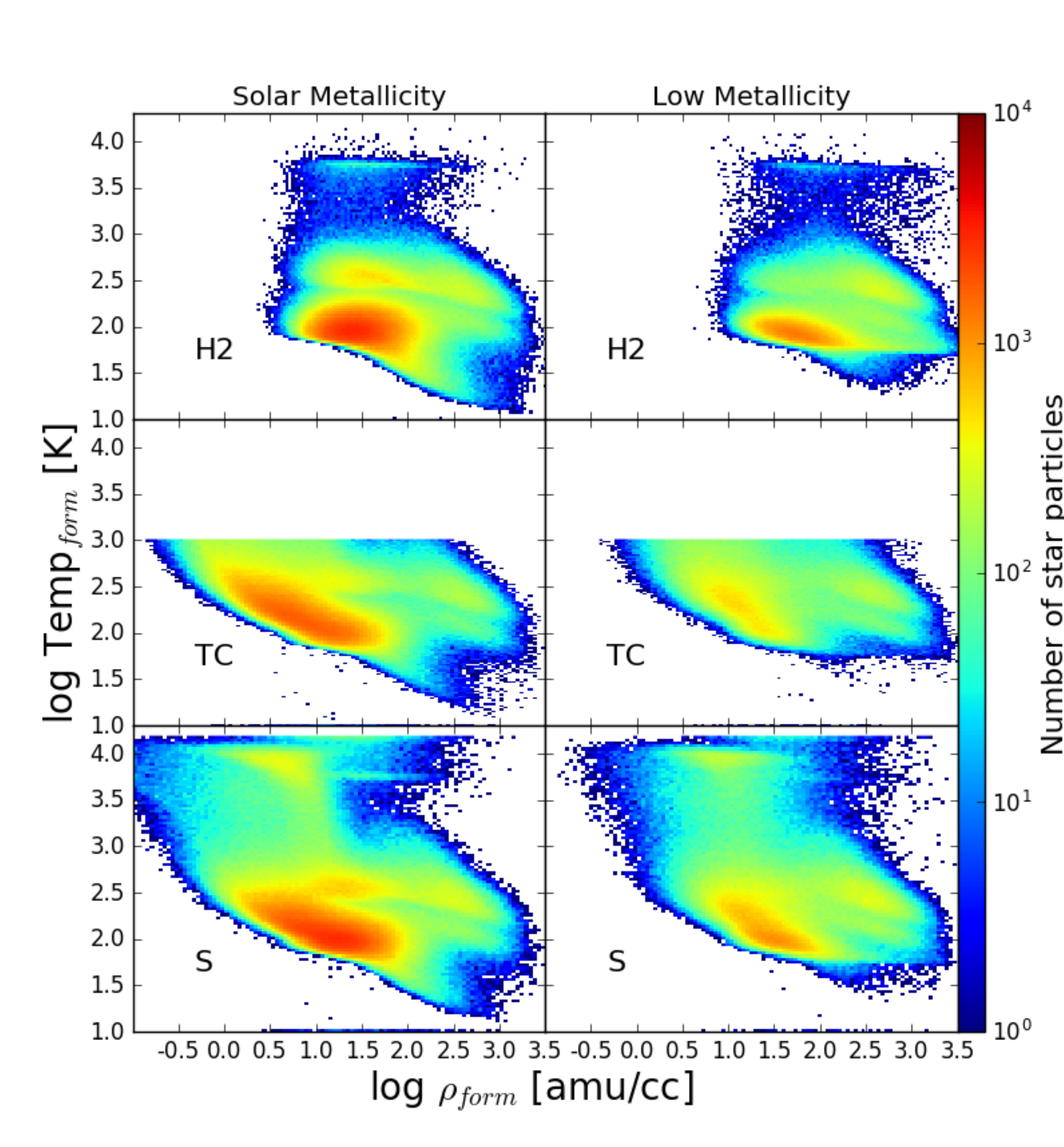}
\end{center}
\vspace{-10pt}
\caption{Phase diagrams of star-forming gas for the solar (left column) and sub-solar metallicity (right column) simulations with H2 (top panels), TC (middle panels) and S (bottom panels) star formation criteria. 
While star formation occurred only in cold, dense gas when linked to the H$_2$ abundance, the shielding and temperature capped models resulted in star formation at higher temperatures and slightly lower densities. There is also a noticeable metallicity effect in which star formation occurred at slightly higher density in the sub-solar metallicity simulations.}
\label{fig:7}
\end{figure}

Narrowing our focus to the star-forming gas, Figure~\ref{fig:7} shows the temperatures and densities at which star formation occurred for all star particles in the solar and sub-solar metallicity simulations. 
In the H2 simulation, no star formation occurred at densities below 1.0 atomic mass unit cm$^{-3}$, and little occurred at temperatures above 1000 K, despite the lack of an explicit temperature cap. 
In the S simulation, however, star formation occurred at densities as low as 0.1 atomic mass units cm$^{-3}$ and temperatures as high as 15000 K. 
While explicitly limiting star formation to gas with temperatures below 1000 K, the TC model also allows for star formation at densities lower than those in the H2 model and similar to the S model. 
When comparing across metallicity, star formation occurred at slightly higher densities in the sub-solar metallicity simulations than the solar metallicity simulations, regardless of the star formation criteria.
In the sub-solar metallicity gas, higher densities were required for the gas to cool, shield, or to maintain H$_2$.

The S simulation resulted in star formation in gas with densities below $1$ amu cm$^{-3}$ and temperatures above $10^3$ K. 
{ Much of this material is low-density gas in thermal equilibrium that is continuing to cool. Those gas particles that are both higher density and hot include gas that, while clearly able to form stars, is not currently in thermal equilibrium because it recently experienced an injection of energy from nearby supernovae. Because of the use of a delayed-cooling feedback model, these particles may currently or until recently have had their cooling disabled.

Although we would not expect significant amounts of star formation at such low densities and high temperatures in reality, the densities and temperatures of gas particles in the simulation should be interpreted as average values for the entire gas cloud.
Therefore, given the limited resolution of the simulations, these low-density, high-temperature star forming gas particles represent clouds of gas that may contain much denser regions.
Finally, the star formation in the shielding simulation was still concentrated at higher densities and lower temperatures than the temperature capped simulation, indicating that although the S model does allow for some low-density star formation, most stars still form out of cold, dense gas as would be expected from observations.  

\begin{figure}
\begin{center}
\includegraphics[width=0.5\textwidth]{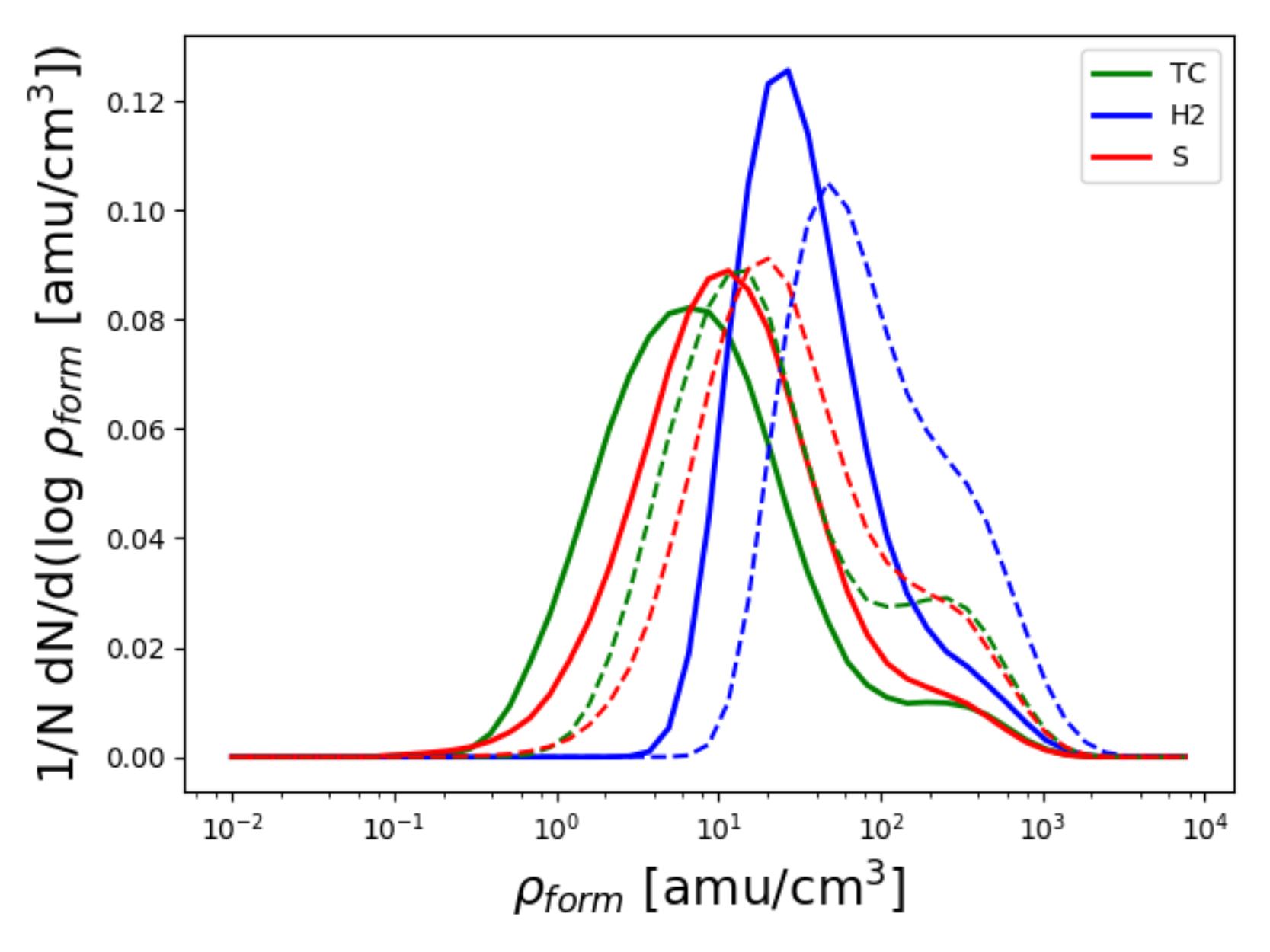}
\end{center} 
\caption{Normalized histograms of the density of star-forming gas. Solid lines indicate solar metallicity simulations, while dashed lines indicate sub-solar metallicity simulations. The H2 model (blue) required the highest average densities for star formation, followed by the S model (red), and then the TC model (green). At lower metallicity, lower shielding and/or lower cooling rates cause star formation to occur at higher densities for all models.}
\label{fig:8}
\end{figure}

We further compare the densities at which star formation occurred in Figure~\ref{fig:8}.
As seen in Figure~\ref{fig:7}, both star formation model and metallicity had a strong effect on the density of the star forming gas.
As predicted, star formation occurred at higher densities in the sub-solar metallicity simulations than the solar metallicity simulations for all models. 
The offset between the peaks of the histograms for the solar and sub-solar simulations was similar regardless of the star formation recipe.
As previously seen, the H2 model required the highest gas densities for star formation, followed by the S model, and then the TC model.

A clear difference between the shielding-based and H$_2$-based star formation models is the reduced density of the star forming gas in the former case.
This difference happens in spite of the close causal link between H$_2$ and dust shielding, and is due to the discrepancy between the chemical equilibrium timescale and the gas collapse timescale in some environments. 
Following the derivation in \citet{Krumholz2012STARGAS}, we can articulate the difference in time scales as follows.
Molecular hydrogen formation occurs on the surface of dust grains, at a rate per H atom of $\bar{n}_H RC$ where $C = \langle n_H^2 \rangle / \bar{n}_H^2$ is a clumping factor and $R \approx 3 \times 10^{-17} Z$ cm$^3$ s$^{-1}$ is the rate coefficient for H$_2$ formation on grain surfaces \citep{Wolfire2008ChemicalFormation}. Then the timescale for conversion of hydrogen from atomic to molecular is: 
\begin{equation}
t_{chem} = \frac{1}{\bar{n}_H R C}.
\end{equation}
The timescale over which gravitationally unstable gas will collapse and begin to form stars is the free-fall time: 
\begin{equation}
t_{ff} = \sqrt{\frac{3\pi}{32G\bar{n}_H \mu_H m_H}}
\end{equation}
where $\mu_H \approx 1.4$ is the mean mass per H nucleus in units of the hydrogen mass $m_H$. The ratio of these timescales is 
\begin{equation}
\frac{t_{chem}}{t_{ff}} = 24Z^{-1}C^{-1}n_o^{-1/2}
\end{equation}
where $n_o = \bar{n}_H/$1  cm$^{-3}$ \citep{Krumholz2012STARGAS} and C = 10 for these simulations.
At low metallicities, the chemical equilibrium timescale can be longer than the free-fall timescale.
For example, for metallicity equal to 0.1 Z$_{\odot}$, $t_{chem} > t_{ff}$ for number densities less than 580 cm$^{-3}$.
This difference in time scales suggests that, assuming H$_2$ is not required for star formation, a cloud of gas could begin to form stars before a substantial amount of H$_2$ is formed.
In our implementation, this resulted in gas forming stars at lower density in the S model than in the H2 model.
In the latter situation, the gas had to form sufficient H$_2$ for star formation to proceed, thereby causing the gas to collapse to higher densities first.
Therefore, the low-density star formation in the shielding model suggests one possible model for the small amounts of star formation that do occur in environments that have low metallicities and low average surface densities, such as dwarf galaxies and the outer disks of spiral galaxies.

\subsection{Star Formation in Cosmological Dwarfs}
\label{sec:results_cosmo}
We examine the impact of the shielding-based star formation prescription on the evolution of a dwarf galaxy.
These simulations were chosen to highlight any differences that might occur between shielding- and H$_2$-based star formation in the low-metallicity, low-surface density environments of dwarf galaxies.
In particular, we investigated the effect of the star formation models on the history and distribution of star formation as well as the resulting galaxy morphology.

To assess the differences in star formation among the various models, we first compared the star formation in our simulated galaxies to the observed redshift zero resolved Kennicutt-Schmidt relation (Figure~\ref{fig:9}). 
To replicate the observations, we calculated surface densities of HI and H$_2$ from mock THINGS \citep{Walter2008THINGS:SURVEY} observations generated at the resolution and sensitivity for a 5 Mpc-distant dwarf galaxy. 
We generated data in 128 velocity channels with 1.3 km s$^{-1}$ bins and a 1.5-arcsec pixel size for the galaxy oriented at a 45\degree angle. We then spatially smoothed the data cubes using a Gaussian beam with a full width at half-maximum of 10 arcsec $\times$ 10 arcsec. Finally, we made a sensitivity cut and discarded all emission from cells below a $2\sigma$ noise limit, in which $\sigma =$ 0.65 mJy beam$^{-1}$.
Star formation rates were calculated using the simulated far UV and 24 $\mu$m emission generated by the radiative transfer code, {\sc sunrise} \citep{Jonsson2006SUNRISE:Geometries}.
Surface densities of gas and star formation were then calculated for 750$^2$ pc$^2$ grid squares.

Both the S and H2 model produce very similar results to each other and to the observed trend for local galaxies \citep{Bigiel2010EXTREMELYGALAXIES}.
Both galaxies have elevated surface densities and star formation rates at their center, resulting in a data point above the observations.
The scatter in the simulated data for locations beyond $r_{25}$ is also somewhat higher than in the observed case.
Otherwise, however, the data from the simulated galaxies follows the same trend as observed galaxies.
While the H2 model results in slightly more high-surface density data points than the S model, both models produce very similar relations to each other.
This similarity is in spite of the fact that the S model results in lower density gas particles forming stars.
Apparently, when the surface densities are averaged over a 750$^2$ pc$^2$ grid, the comparatively high densities of the star forming gas in the simulation with the H2 model were diluted by the surrounding interstellar media.
Similar results were found in \citet{Christensen2012ImplementingFormation}, where a density threshold-based star formation law produced a similar Kennicutt-Schmidt relation to the H$_2$-based star formation law.
We therefore conclude that despite the differences between the shielding and H$_2$-based star formation models, both are consistent with the resolved Kennicutt-Schmidt relation.

\begin{figure}[t]
\begin{center}
\includegraphics[width=0.5\textwidth]{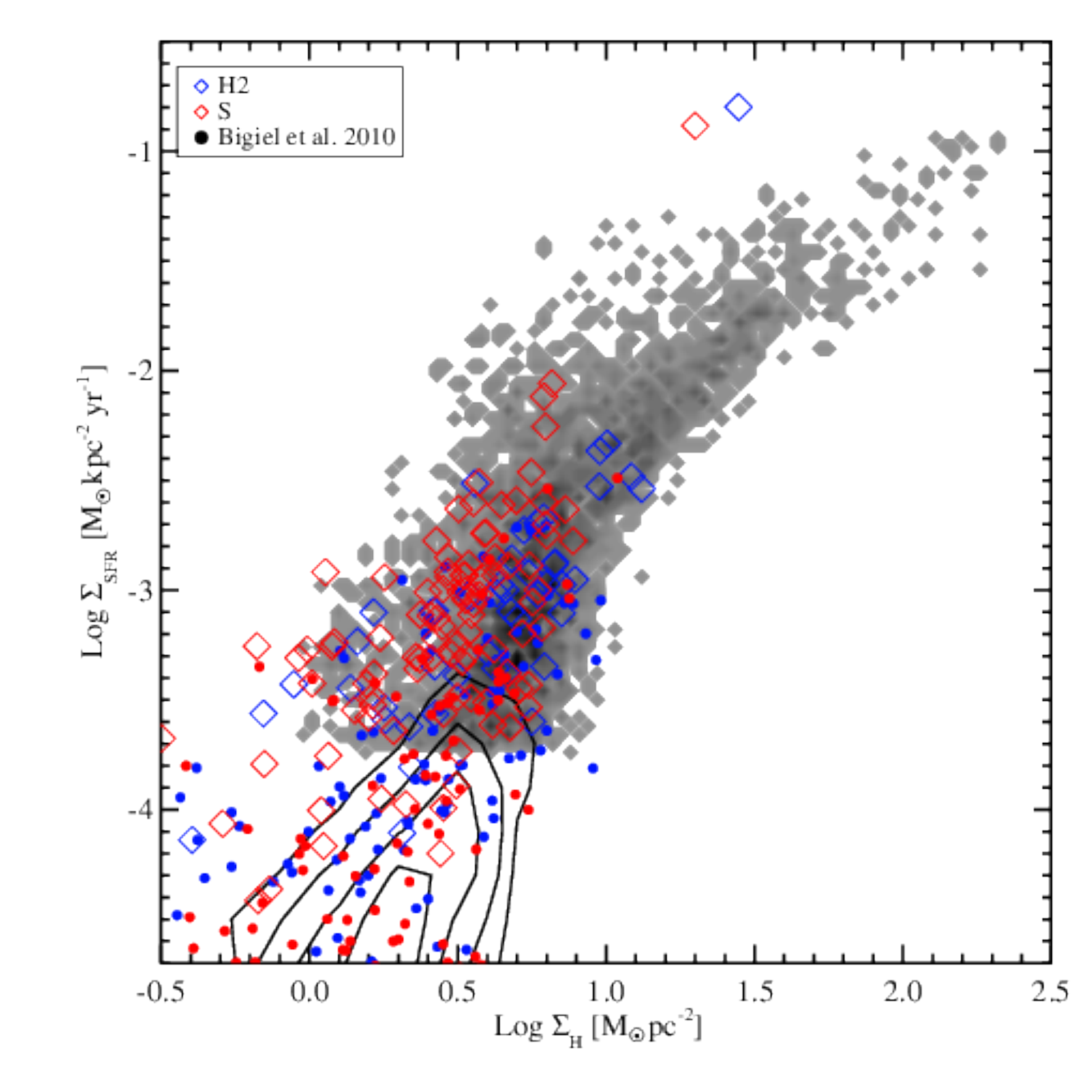}
\end{center}
\caption{
The redshift zero resolved Kennicutt-Schmidt relation for cosmological simulations using the S and H2 models. {\bf Blue points are data from the dwarf galaxy simulated with the H2 model, and red points are from the simulation using the S model. Diamonds show data from within $r_{25}$, while circles designate data between $r_{25}$ and $2 r_{25}$.} In both cases, the simulated data was determined using mock-HI, H$_2$, far UV and 24 $\mu$m observations.
 The greyscale represents data within $r_{25}$ from \cite{Bigiel2010EXTREMELYGALAXIES} for local galaxies.
Black contours enclose 90, 75, 50, and 25\% of data points with detectable star formation rates located between $r_{25}$ and $2 r_{25}$} for the sample from \cite{Bigiel2010EXTREMELYGALAXIES}. Only simulated and observed data with star formation rates above the observable lower-limits are shown.
\label{fig:9}
\end{figure}

One environment where shielding-based star formation might occur is in the low-metallicity outskirts of galaxies \citep{Ferguson1998DiscoveryGalaxies,Bigiel2010EXTREMELYGALAXIES,Goddard2010OnRadii} 
In these regions, star formation is observed to occur despite very low levels of H$_2$, and it is possible that the initial gravitational collapse for star formation might occur prior to H$_2$ formation.
We therefore compare the ISM, stellar, and star formation rate surface densities as a function of radius (Figure~\ref{fig:10}). Both models produce similar gas profiles, though the H2 model does produce a higher surface density of H$_2$.
Likewise, the stellar profiles are almost identical.
However, we do find that the star formation rate surface density in the S model is slightly higher than in the H2 model, particularly in the outer disk, despite the lower H$_2$ surface densities. 

\begin{figure}[t]
\begin{center}
\includegraphics[width=0.5\textwidth]{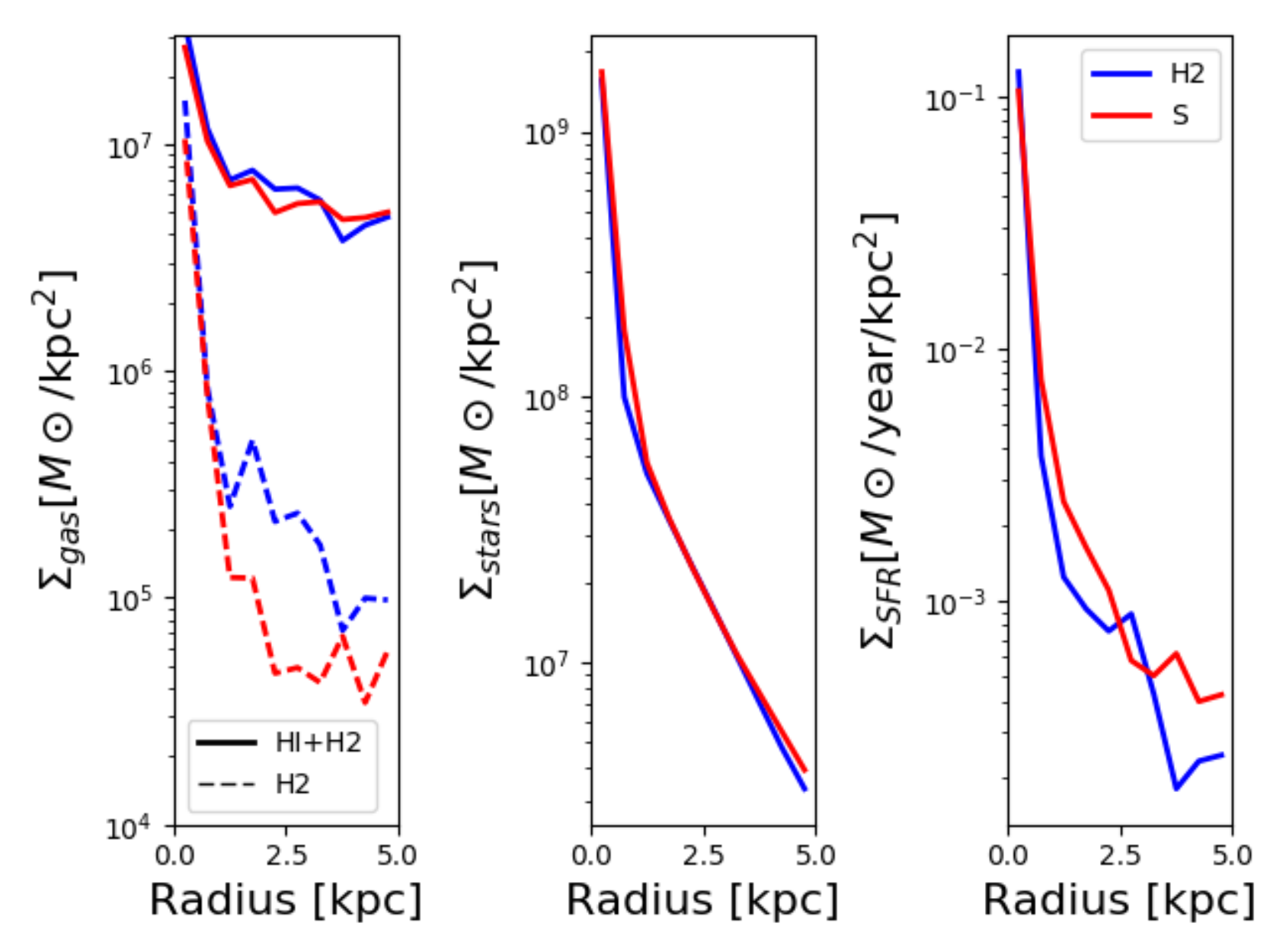}
\end{center}
\caption{Radial profiles of surface density of HI and H$_2$ gas (left), stellar surface density (center), and surface density of star formation rate (right) for the cosmological simulations. 
Blue lines show the simulation computed with the H2 star formation model and red lines the one computed with the S model. The star formation rate was calculated for the last 50 Myr of the simulation.
The gas and stellar surface density profiles are similar for the two simulations, while the star formation rate is higher in the outer disk for the S model.}
\label{fig:10}
\end{figure}

Although the average gas surface densities were similar between the simulations produced with the H2 model and the S model, a detailed analysis of the particle densities reveals differences (Figure ~\ref{fig:11}). 
We found that the H2 simulation produced more numerous and higher density clumps of gas than the S simulation. 
These clumps are evident in the figure as spatially localized spikes in the particle densities.
This difference in gas densities follows from the tendency for star formation to occur in lower density gas in the S simulation.
As a result, the galaxy was less able to maintain high density clumps of material.
The similarity in the average surface densities between the models, which was also apparent in the K-S law (Figure~\ref{fig:9}) and gas profile (Figure~\ref{fig:10}), camouflages these small scale differences.

\begin{figure}[t]
\begin{center}
\includegraphics[width=0.5\textwidth]{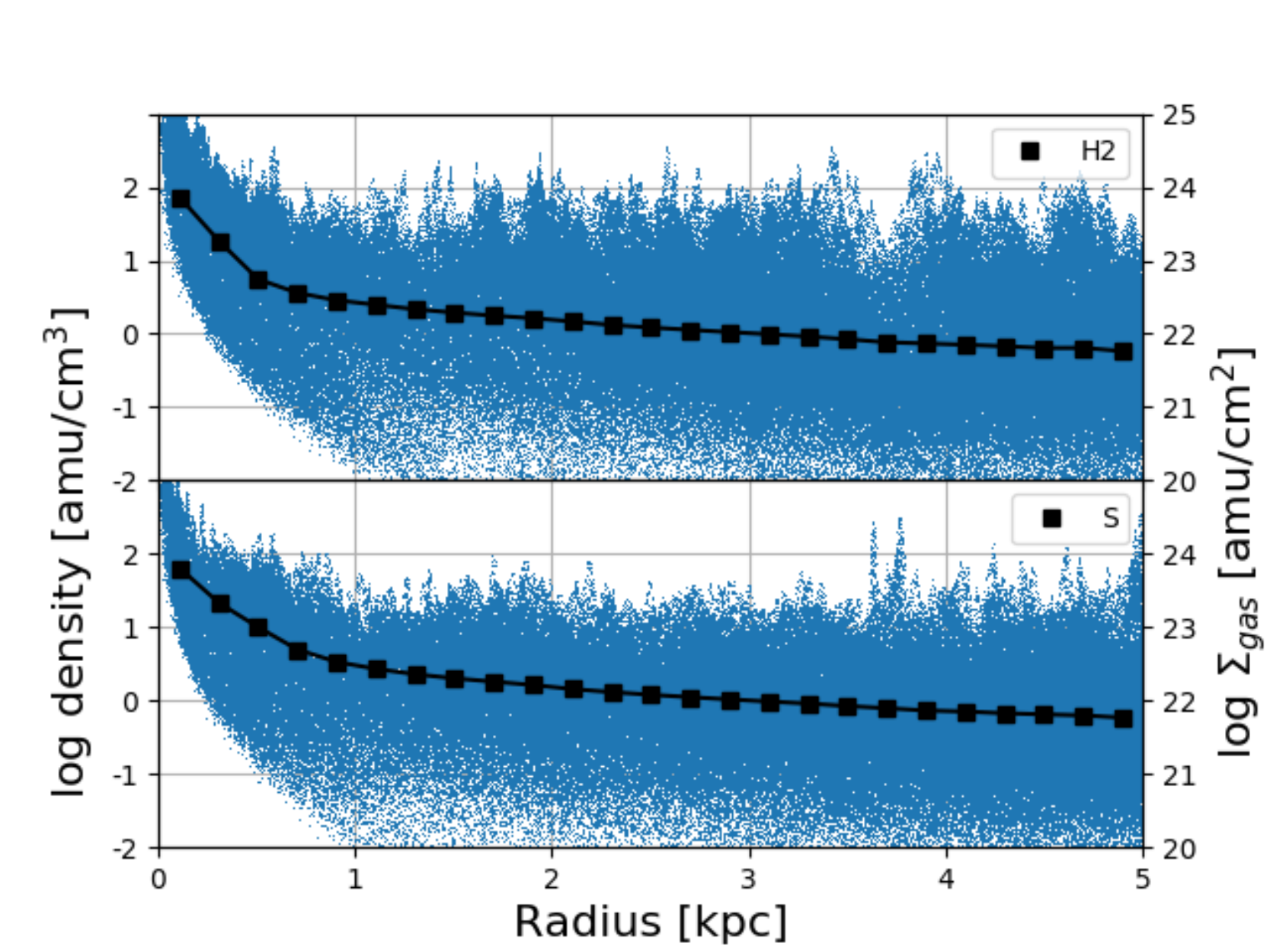}
\end{center}
\caption{Density and surface density of gas in the inner disk for the cosmological H2 and S simulations. Blue points represent individual gas particle densities, while the connected black points show the average surface density calculated in radial bins. 
The top panel shows the result of the H2 model and the bottom panel the result of the S model.
Though the overall surface density is consistent between the two models, the H2 simulation has more and higher density clumps of gas, visible in the plot as spikes in the particle densities.}
\label{fig:11}
\end{figure}

\begin{figure}
\begin{center}
\includegraphics[width=0.5\textwidth]{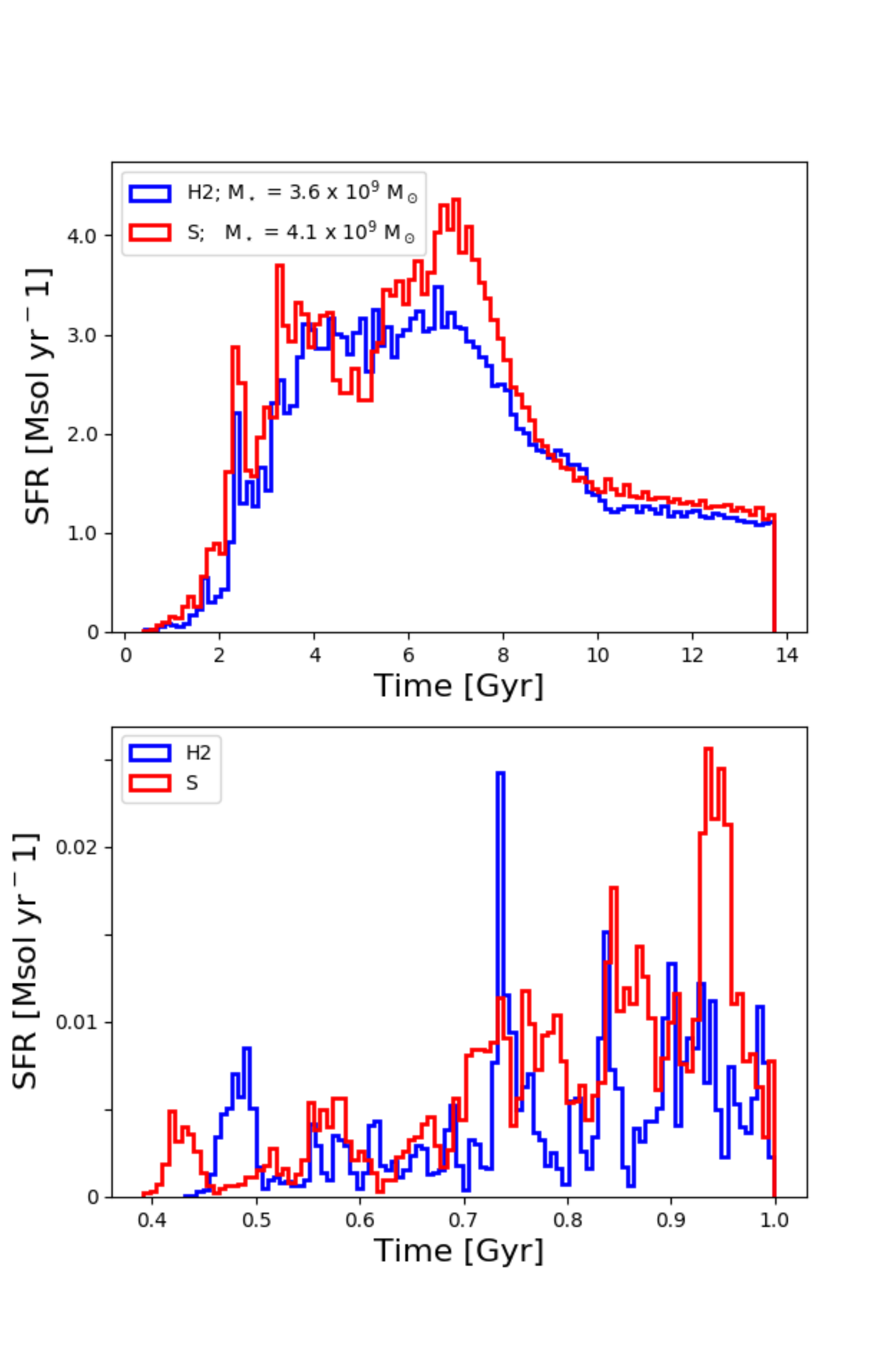}
\end{center} 
\caption{Star formation rate against time since the beginning of the simulation for cosmological dwarf galaxy simulations computed with the H2 (blue) and S (red) star formation. The top panel shows the history over the entire simulation, while the lower panel shows only the first Gyr. The value M$_\star$ listed in the legend of the top panel is the total stellar mass formed over the course of each simulation. Both models produce similar star formation histories, though the S model leads to a slightly greater amount of star formation. Star formation also began earlier in the galaxy's history for the S model. Both models created periodic bursts of star formation at early times, as seen in the lower panel, but not at later times.}
\label{fig:12}
\end{figure}

Finally, we compare the star formation across the entire history of the galaxy. 
The total stellar masses formed over the course of the simulations were $3.6\times 10^9$ M$_\odot$ for the H2 model and $4.1\times 10^9$ M$_\odot$ for the S model, indicating that the star formation recipe had a significant effect on the total mass of stars produced. 
This difference in stellar mass could have been due to either or both of the following possible causes. First, star formation may have been allowed in a wider range of environments in the S simulation. From Figures \ref{fig:7} and \ref{fig:8}, it is clear that the S model expands the range of gas particles capable of star formation, supporting this possibility. Secondly, changes to the star formation environments caused by the different models could have had other indirect effects; in particular, the efficiency of the feedback recipe may have been affected. This possibility is supported by previous research with similar simulations. 
For example, \citet{Governato2010BulgelessOutflows} and  \citet{Christensen2012ImplementingFormation} showed that the blastwave feedback recipe is most effective at regulating star formation when star formation is confined to high-density clumps.
It is likely that both explanations play a role here in the total mass formed.
It is also important to understand that the amount of feedback energy injected per SNe (dESN) is a tunable parameter that must be set for each combination of star formation parameters.
While we used the same value of dESN for both simulations to create a fair comparison, when using either star formation model in other research, dESN would be set to ensure appropriate z = 0 stellar masses. 
 
Looking in more detail at the star formation histories (Figure~\ref{fig:12}), it is clear that the S model begins star formation earlier, resulting in greater amounts of star formation within the first 2 Gyrs.
During this early time period, the metallicity of the ISM is extremely low. 
Low metallicity not only increases the surface densities at which dust shielding occurs, it also decreases the rate of H$_2$ formation on dust grains.
As a result, while low metallicity limits star formation in both the S and H2 models, it has a stronger effect with the H2 model.

After 2 Gyrs, the star formation histories have a similar shape, though the star formation rate is consistently slightly higher in the S simulation. 
Both the H2 and S recipe produce bursty star formation histories at high redshift and smoother star formation histories at low redshift. 
Bursty star formation histories with periods of 150-200 Myr are characteristic of present-day dwarf galaxies \citep{Weisz2012MODELINGGALAXIES}.
To resolve bursts on these timescales, we examine the star formation history over a 1 Gyr time period (Figure~\ref{fig:12}). 
Bursts of star formation lasting 50-100 Myr appear in both the H2 and S simulations during the first Gyr but are not present in either during the latter part of the simulation, even when examining 1 kpc$^2$ subsections of the disk.
Continually bursty star formation histories for dwarf galaxies are characteristic of simulations with efficient supernova feedback and high star formation density thresholds \citep{ Governato2010BulgelessOutflows, Hopkins2014GalaxiesFormation, Shen2014THEPOLLUTERS,Sparre2017StarburstsGalaxies}.
Since both the H2 and S simulations have these qualities, it is somewhat surprising that neither maintain bursty star formation across the entire history of the galaxies.
Nevertheless, the similar shapes of the star formation histories indicate that both the H2 and S star formation models are operating in a similar fashion.

\section{Discussion and Conclusions}
In this study we developed and analyzed an implementation for dust shielding-based star formation in galaxy evolution simulations.
We examined the effect of tying star formation probability to the amount of dust shielding, rather than the fraction of molecular hydrogen, finding that the former resulted in star formation in lower-density gas while still being consistent with the Kennicutt-Schmidt relation.
We summarize our implementation and these effects below. 

The shielding-based star formation model consisted of a stochastic sub-grid model in which the probability of a gas particle forming a star particle was a function of the amount of dust shielding.
This dependency on dust shielding replicates the theoretical model in which only shielded gas is able to achieve the low temperatures necessary for gravitational collapse.
As part of the model development, different prescriptions were used for calculating the column length associated with a given gas particle. 
In agreement with \cite{Safranek-Shrader2017ChemistryDiscs}, we found that a prescription in which the column length is the Jeans length calculated with a 40 K temperature cap produced the best agreement with observations.
With this prescription, we were able to replicate the observed transition from HI to H$_2$ as a function of the HI + H$_2$ surface densities calculated for the disk during post-processing.

We compared our shielding-based star formation model to a molecular hydrogen-based star formation model and a temperature ceiling model using simulations of isolated disk galaxies. 
The primary differences between the models were the densities and temperatures of the star forming gas.
The shielding-based model resulted in star formation at higher temperatures and lower densities than the H$_2$-based model, while still requiring higher densities on average than a temperature ceiling recipe. 
One consequence of the different star forming gas densities was the presence of denser gas in the simulation using the H2 model, which also lead to more pronounced spiral structure.

The denser star forming gas in the H2 model can be explained by the time delay introduced by requiring the gas to form H$_2$ prior to star formation.
During the time that the gas is forming H$_2$, it also may gravitationally collapse and reach higher densities before spawning a star particle.
In models that do not rely on an explicit link between H$_2$ and star formation, the discrepancy between free-fall timescales and molecular formation timescales allows star formation to begin earlier while the gas particle is at a lower density. 

We further compared the shielding and H$_2$ star formation models by examining the evolution of cosmological simulations of dwarf galaxies.
Both the shielding and H2 models produced similar fits to the observed Kennicutt-Schmidt relation when resolved on 750 pc scales.
At this resolution, differences in the densities of the star forming gas were washed out, and both recipes provide realistic results.
Despite similar gas surface density profiles, the star formation rate in the outer disk was slightly higher for the shielding simulation than the H2 simulation.
This difference in star formation at large radius is likely the result of shielding-based star formation being able occur at lower densities.
However, the resulting stellar profiles were essentially the same for both models.
Likewise, both star formation models produced similarly shaped star formation histories with similar levels of burstiness.
The exception to this similarity is that star formation is able to begin slightly earlier when using the S model, because extremely low metallicity has a weaker effect on the presence of dust shielding than the amount of H$_2$. The star formation rate continued to be higher when using the S model for most of the duration of the simulation. 

The strongest difference the star formation models produced in the cosmological simulations was in the total mass of stars formed.
The shielding model resulted in approximately ten percent more stellar mass formed over the history of the universe than the H2 model. The additional stellar mass appears to be driven by a higher star formation rate in the outer disk.
This difference appears to be caused not by differences in the amount of star formation per cold gas, as seen in the Kennicutt-Schmidt relation, but by the existence of greater amounts of cold gas in the H2 simulation.
We attribute this differences to a decrease in the effective efficiency of the supernova feedback model when star formation is allowed to take place in lower density gas. 
Similar behavior was observed in \citet{Christensen2012ImplementingFormation}, where one of the main impacts of using an H$_2$-based star formation recipe was the enhanced feedback efficiency due to star formation happening in higher density gas.
The slightly lower relative efficiency of feedback observed when using the shielding model can be modulated by increasing the amount of energy released per supernovae mass, $dESN$.
We therefore stress that while both dust shielding and H$_2$-based star formation models produce realistic behavior (e.g. matching the Kennicutt-Schmidt relation), the tuning of $dESN$ and other star formation or feedback models must be model specific.

While we found that the dust shielding-based model enable star formation in lower density gas than the H$_2$-based model, there were only slight differences in the observable properties of the resulting galaxies.
It is likely that these effects would be stronger in simulations that focused on the extremely early Universe or that resolved the internal structure of star forming clouds.
On the resolution and time scales of cosmological galaxy formation simulations, though, the different models do not produce markedly different observable characteristics.
Therefore, we conclude that both the dust shielding-based and H$_2$-based star formation models are viable and realistic in the context of galaxy formation simulations.

\acknowledgments
\section{Acknowledgments}
The authors would like to thank the anonymous referee for insightful comments.
LB acknowledges support through the Grinnell College Mentored Advanced Project program.
The cosmological simulations were computed at NASA AMES.
Analysis of the simulations was undertaken using the Python analysis package for astrophysical N-body and Smooth Particle Hydrodynamics simulations, {\sc pynbody} \citep{Pontzen2013Pynbody:Python}.

\bibliographystyle{yahapj}
\bibliography{Mendeley_Shielding_2017}

\end{document}